\documentclass{article}
\usepackage[utf8]{inputenc}
\usepackage{libertine}
\usepackage{amsthm,amsmath,amssymb,amsfonts}
\usepackage{mathtools}
\usepackage{mathrsfs}
\usepackage{bbm} 

\def\Natural{\mathbb{N}}

\def\Real{\mathbb{R}}
\def\sigmaB{\mathcal{B}}
\def\sigmaF{\mathcal{F}}

\def\Prob{\mathbf{P}}
\def\ind{\mathbbm{1}}
\newcommand\Expect[1]{\mathbf{E}  #1 }
\newcommand\E\Expect
\def\mix{\mathop\mathrm{mix}\limits}
\DeclareMathOperator{\ES}{ES}
\def\Set#1#2{\left\{{#1}\ \colon\ {#2} \right\}}

\newcommand{\defeq}{\coloneqq}
\newcommand{\eqd}{\overset{\mathrm{d}}{=}}
\def\toX#1{\overset{#1}{\to}}
\def\tok{\toX{\kappa}}
\def\fosd{\stackrel{\textnormal{1\,sd}}{\succcurlyeq}}
\def\sosd{\stackrel{\textnormal{2\,sd}}{\succcurlyeq}}
\def\div#1{\stackrel{\textnormal{div}_{#1}}{\succcurlyeq}}
\def\divk#1{\stackrel{\textnormal{div}_{#1}(\kappa)}{\succcurlyeq}}
\def\k{\kappa}
\def\Lip#1{\mathrm{Lip}_{ #1 }}
\def\Law{\mathscr{L}}

\newtheoremstyle{my_def_style} 
{5pt} 
{3pt} 
{} 
{} 
{\scshape\bfseries} 
{.} 
{.5em} 
{} 

\newtheoremstyle{my_thm_style} 
{5pt} 
{3pt} 
{\itshape} 
{} 
{\scshape\bfseries} 
{.} 
{.5em} 
{} 

\theoremstyle{my_def_style}
\newtheorem{my_def}{Definition}

\theoremstyle{my_thm_style}
\newtheorem{my_lemma}{Lemma}
\newtheorem{my_thm}{Theorem}
\newtheorem{my_cor}{Corollary}

\newenvironment{my_prf}[1][\proofname]{\par\noindent\pushQED{\qed}\scshape\bfseries#1. \normalfont\ignorespaces}{\popQED}

\title{On a Stochastic Model of Diversification}
\author{Maria Logvaneva, Mikhail Tselishchev \thanks{Department of Mathematical Statistics, Faculty of Computational Mathematics and Cybernetics, Lomonosov Moscow State University. Email: \textit{mihail.tselishchev(at)gmail(dot)com}.}}
\date{April 4, 2022}

\begin{document}

\maketitle
\thispagestyle{empty}  

\begin{abstract}
We propose a definition of diversification as a binary relationship between financial portfolios. 
According to it, a convex linear combination of several risk positions with some weights is considered to be less risky than the probabilistic mixture of the same risk positions with the same weights.
It turns out to be that the proposed partial ordering coincides with the well-known second order stochastic dominance, but 
allows to take a look at it from another perspective.

\end{abstract}

\section{Introduction}

The principle of diversification postulates that allocating capital between assets reduces investment risk.
Intuitively, a combination of identical (but not necessarily independent) assets is less risky than each individual one.
It the paper we take this idea as a basis for our model of diversification and analyze implications.

In order to give mathematical definitions we will need some basic notation. 
We assume that risk positions are random variables (r.v.s) 
on some probability space $(\Omega,\sigmaF,\Prob)$,
they denote profits (or losses when negative) of financial portfolios 
at the end of the trading period.
The cumulative distribution function (c.d.f.) of risk position~$\xi$ is denoted by
$F_\xi(x) = \Prob(\xi < x)$, $x\in\Real$.

It turns out to be that in case of finite expectations, the proposed model of diversification is tightly related to the preference of \textit{the second order stochastic dominance}:
$$
\int_{-\infty}^a F_\xi(x) \, dx 
\; \le \; \int_{-\infty}^a F_\eta(x) \, dx 
\qquad \text{for every} \quad a\in\Real,
$$
that is denoted by $\xi \sosd \eta$.
Namely, we prove that in case of equal expectations, both relations coincide after closing the diversification dominance in space of probability distributions on $(\Real, \sigmaB)$ with finite first moments w.r.t. Kantorovich metric.

The second order stochastic dominance allows several equivalent representations, in terms of utility functions and also in terms of a widely-used coherent risk measure called \textit{Expected Shortfall} (see~\cite{AcerbiTasche_ES},\cite{AcerbiTasche_OnTheCoherence}).
Expected Shortfall of risk position $X$ (with $\E X^- < \infty$) at level $\alpha\in(0,1]$ is defined as
$$
\ES_\alpha(X) \defeq - \frac{1}{\alpha} \int_0^\alpha q_X(u) \, du,
$$
where
$$
q_X(u) \defeq \inf \Set{x \in \overline\Real}{F_X(x) \ge u}, \quad u\in[0,1],
$$
is a lower-quantile function for distribution of r.v. $X$.
The Expected Shortfall at level~$\alpha$ shows an average loss of $X$ in the worst $\alpha \cdot 100 \%$ scenarios.

\begin{my_thm} \label{th:2sd:criteria}
Let $\xi$ and $\eta$ be r.v.s with finite expectations.
Then the following statements are all equivalent:
\begin{itemize}
\item $\xi \sosd \eta$,
\item $\E{u(\xi)} \ge \E{u(\eta)}$ for all non-decreasing concave functions $u\colon \Real \to \Real$,
\item $\ES_\alpha ( \xi ) \le \ES_\alpha ( \eta )$ for all $\alpha \in (0,1]$.
\end{itemize}
If, in addition, $\E \xi = \E \eta$, then the previous statements are equivalent to
\begin{itemize}
\item $\eta$ is a mean-preserving spread of $\xi$, i.e.
$\eta \eqd \xi + \varepsilon$ for some $\varepsilon$ 
with $\E{(\varepsilon|\xi)} = 0$.
\end{itemize}
\end{my_thm}

\noindent
The proof of Theorem~\ref{th:2sd:criteria} may be found, for instance, 
in~\cite[Th.~2.57 and Cor.~2.61]{FollmerSchied2016}.

\section{Definitions \& Results}

Starting with the notation
$$
\mathcal S_{m-1} \defeq \Set{(x_1,\ldots,x_m) \in \Real^m}{x_i \ge 0 \text{ for all } i=1,\ldots, m \text{, and } \sum_{i=1}^m x_i = 1}
$$
of a standard $(m{-}1)$-simplex, we will give the following definitions.

\begin{my_def} \label{def:div1}
We say that the risk position $\xi$ dominates the risk position $\eta$ in terms of diversification and denote it by
$\xi {\div1} \eta$, if there exist a random vector $X=(X_1,X_2,\ldots, X_m)$ and weights $\beta \in \mathcal S_{m-1}$, such that $X_i \eqd \eta$ for all $i=1\ldots m$ and $\xi \eqd \sum_{i=1}^m \beta_i X_i$.
\end{my_def}

In other words, a convex linear combination of identically distributed (but not necessarily independent) risk positions is better than each individual of them in terms of diversification. This seems to be very intuitive.

Note that such random vector $X$ possibly may not exist on the original probability space if this space is rather poor (e.g. finite), but may exist on some other space. So, one may think of probability distribution on $\Real^m$ with the specified properties rather than of random vector $X$.

One may ask why should we take all $X_i$ having the same distribution?
If we relax this requirement, the question immediately arises: a convex linear combination $\sum_{i=1}^n \beta_i X_i$ of risk positions is better than \textit{what}?
We insist that the answer should be: it is better than a~mixture of the same risk positions with the same weights, which we denote by $\mix_\beta X$. This is a random variable having c.d.f.
$F(x) = \sum_{i=1}^m \beta_i F_{X_i}(x)$.

\begin{my_def} \label{def:div2}
$\xi \div2 \eta$ if there exist a random vector $X=(X_1,X_2,\ldots, X_m)$ and weights $\beta \in \mathcal S_{m-1}$, such that $\xi \eqd \sum_{i=1}^m \beta_i X_i$ and $\eta\eqd\mix_\beta X$.
\end{my_def}

Note that both definitions~\ref{def:div1} and~\ref{def:div2} only compare risk positions with the same expectations: $\xi\div{j}\eta$ implies either $\Expect \xi = \Expect \eta$ or $\Expect{|\eta|} = +\infty$.
Furthermore, it's evident that this is a comparison of probability distributions on $(\Real,\sigmaB)$ rather than a comparison of random variables.

As we show below, there is not much difference between two presented definitions, especially after closing them.

\begin{my_lemma} \label{lem:div_implies_sosd}
$\xi \div1 \eta$ implies $\xi \div2 \eta$, and $\xi \div2 \eta$ implies $\xi \sosd \eta$ for r.v.s with finite expectations.
\end{my_lemma}

\begin{my_prf}
The first implication is trivial.
The second one follows from Theorem~\ref{th:2sd:criteria}, the convexity of $\ES$ w.r.t. risk positions and the concavity of $\ES$ w.r.t. probability distributions, namely
$$
\ES_\alpha\left( \sum_{j=1}^m \beta_j X_j \right)
\le \sum_{j=1}^m \beta_j \ES_\alpha (X_j)
\le \ES_\alpha \left( \mix_\beta X \right).
$$
The proof of the latter can be found in~\cite{Tselishchev2019} and 
also\footnote{
Special thanks to Ruodu Wang who pointed out in a private email that the property of mixture-concavity of $\ES$ has been known for a long time.
}
in~\cite{Wang2019}.
\end{my_prf}

We need to recall the classical Farkas' lemma to prove next results.

\newtheorem*{fl}{Farkas' lemma}
\begin{fl}
Let $B\in\Real^{n\times m}$ and $a \in \Real^n$. Exactly one of the following alternatives holds true:
\begin{itemize}
\item system $Bx=a$ has a solution $x\in\Real^m_+$,
\item there exists $y\in\Real^n$, such that $a^T y < 0$ and $B^T y \ge 0$.
\end{itemize}
\end{fl}

\begin{my_lemma} \label{lem:simple:sosd_implies_div1}
Let $\xi$ and $\eta$ be simple r.v.s taking their values with rational probabilities. If\;~$\E{\xi} = \E{\eta}$ and $\xi\sosd \eta$, then $\xi \div1 \eta$.
\end{my_lemma}

\begin{my_prf}
Without loss of generality one may assume that
$$
\xi = \sum_{i=1}^n a_i \ind_{A_i}
\qquad \text{and} \qquad
\eta = \sum_{i=1}^n b_i \ind_{A_i},
$$
where $a_1\le a_2 \le \ldots \le a_n$, $b_1\le b_2 \le \ldots \le b_n$ and $\bigsqcup\limits_{i=1}^n A_i = \Omega$, $\Prob(A_i) = \frac1n$.
Since both relations $\sosd$  and $\div1$ are translation invariant, one also may assume that $a_1 > 0$, $b_1 > 0$.
The equality of expectations $\E\xi = \E\eta$ means that 
\begin{equation} \label{lem:prf:means_equal}
\sum_{i=1}^n a_i = \sum_{i=1}^n b_i.
\end{equation}
Due to Theorem~\ref{th:2sd:criteria}, relation $\xi\sosd \eta$ means that
\begin{equation} \label{lem:prf:2sd}
\sum_{i=1}^j a_i \ge \sum_{i=1}^j b_i
\qquad \text{for all} \quad j=1\ldots n.
\end{equation}
Define $X_k = \sum\limits_{i=1}^n b_{\sigma_k[i]} \ind_{A_i}$, where $\sigma_k$ is the $k$-th permutation of $(1,.., n)$, $k=1\ldots n!$.
Obviously, all $X_k \eqd \eta$, and in order to show $\xi \div1 \eta$ we are going to prove that there exists $\lambda\in\mathcal S_{n!-1}$, such that 
$\xi = \sum\limits_{k=1}^{n!} \lambda_k X_k$, or, in other words,
\begin{equation} \label{lem:prf:system}
B \lambda = a,
\end{equation}
where $a=(a_1,\ldots,a_n)^T$ and the columns of matrix $B\in\Real^{n\times n!}$ are all possible permutations of vector $b=(b_1,\ldots,b_n)^T$.
Suppose~\eqref{lem:prf:system} has no positive solutions.
Then, by Farkas' lemma, there should exist $y\in\Real^n$ such that $a^T y < 0$ and $B^T y \ge 0$.
The latter means that $b_\sigma^T y \ge 0$ for any permutation $\sigma$.
Note that one can take~$y$ ordered ($y_1 \ge \ldots \ge y_n$), since 
$$
0 > a^Ty = \sum_{i=1}^n a_i y_i \ge \sum_{i=1}^n a_i y_{(n-i+1)}
$$
due to $0 < a_1 \le a_2 \le \ldots \le a_n$.
By using algebraic transformations, notation $y_{n+1} = 0$, equality~\eqref{lem:prf:means_equal} and inequalities~\eqref{lem:prf:2sd}, we get
\begin{multline*}
0 > a^T y 
= \sum_{i=1}^n a_i\left( \sum_{j=i}^n (y_j - y_{j+1}) \right)
= \sum_{j=1}^n\left( (y_j - y_{j+1}) \sum_{i=1}^j a_i\right) \ge 
\\
\ge \sum_{j=1}^n\left( (y_j - y_{j+1}) \sum_{i=1}^j b_i\right)
= \sum_{i=1}^n b_i\left( \sum_{j=i}^n (y_j - y_{j+1}) \right) = b^T y \ge 0,
\end{multline*}
that leads to contradiction. Thus, system~\eqref{lem:prf:system} has a solution $\lambda \in \Real^{n!}_+$. 
\\Finally, to see that $\lambda \in \mathcal S_{n!-1}$, one has to sum all the equations in system~\eqref{lem:prf:system} and use~\eqref{lem:prf:means_equal}.
\end{my_prf}

\begin{my_lemma}
Both relations $\div1$ and $\div2$ are not closed.
\end{my_lemma}

\begin{my_prf}
Let $\eta_1,\eta_2,\ldots \stackrel{\mathrm{i.i.d.}}{\sim} \mathrm{Exp}(1)$. 
Denote $\xi_n \defeq \frac{1}{n}\sum_{i=1}^n \eta_i$. 
Clearly, $\xi_n {\div1} \eta_1$ for all $n\in\Natural$.
By the law of large numbers, $\xi_n$ converges to $1$ as $n\to\infty$, both almost sure and in~$L^1$, so one may suspect that $1 \div2 \eta_1$.
If so, there exist a random vector $X=(X_1,\ldots,X_m)$ and weights $\beta \in \mathcal S_{m-1}$, such that
$$
\sum_{i=1}^m \beta_i X_i = 1 \text{ \ a.s. \ \quad and \quad}
\mix_\beta X \sim \mathrm{Exp}(1).
$$
Assuming all $\beta_i > 0$, the latter implies $\Prob(X_i > 0) = 1$ for all $i=1\ldots m$ and $\Prob(X_k > a) > 0$ for some $k$ and all $a > 0$. Hence,
$$
\Prob\left(\sum_{i=1}^m \beta_i X_i > 1\right) \ge \Prob\left(X_k > \frac{1}{\beta_k}\right) > 0,
$$
that gives a contradiction.
\end{my_prf}

Previous two lemmas suggest to perform a closure of the proposed relations in some metric space. We will use the space of all distributions on $(\Real,\sigmaB)$ with
finite first moments endowed with Kantorovich metric $\k$ that has several equivalent representations (see, e.g.,~\cite[Sect.~3.2]{bogachev2018}):
\begin{multline} \label{def:kantorovich_metric}
\k\left(\xi,\eta\right) 
= \sup_{h \in \Lip1} \left| \int_\Real h\,dF_\xi - \int_\Real h\,dF_\eta \right|
= \min_{\Law(X,Y)\colon X\eqd \xi,\,Y\eqd \eta}\Expect\,|X-Y|
=\\= \int_0^1 \left| q_\xi(u) - q_\eta(u) \right| \, du
= \int_{-\infty}^{+\infty} \left| F_\xi(x) - F_\eta(x) \right| \, dx.
\end{multline}
where
$\Lip1 = \Set{h\colon \Real\to\Real}{ |h(x) - h(y)| \le \,|x-y| \quad \forall x,y \in\Real }$.

\begin{my_def} 
We say that the risk position $\xi$ dominates the risk position $\eta$ 
in terms of the closure of the relation $\div{j}$ in metric $\k$,
and denote it by $\xi {\divk{j}} \eta$, if there exist two sequences of r.v.s $\xi_n$ and $\eta_n$, such that $\xi_n \tok \xi$, $\eta_n \tok \eta$ and $\xi_n {\div{j}} \eta_n$ for all $n\in\Natural$.
\end{my_def}

As before, it may be better to think of the probability distributions on $(\Real,\sigmaB)$ rather than of r.v.s.

In the definition above one may take any other metric, 
but we mainly focus on the closure in Kantorovich metric, 
since it provides nice features.

To prove the main result, we will need another auxiliary lemma.

\begin{my_lemma} \label{lem:auxilary}
Let $\xi$ and $\eta$ be simple r.v.s taking their values with rational probabilities. 
Then there exist simple non-negative r.v.s $\delta$ and $\gamma$ taking their values with rational probabilities, such that
$\xi+\delta \div1 \eta + \gamma$ and
\begin{equation} \label{lem:eq:mean_delta}
\E{\delta} = \sup_{\alpha\in(0,1]} \alpha \cdot\left(\big.\ES_\alpha(\xi) - \ES_\alpha(\eta)\right), 
\qquad
\E{\gamma} = \E{\xi}  -\E{\eta} + \E{\delta}.
\end{equation}
\end{my_lemma}

\begin{my_prf}
Again, as in the proof of Lemma~\ref{lem:simple:sosd_implies_div1},
one may assume without loss of generality that
$$
\xi = \sum_{i=1}^n x_i \ind_{A_i}
\qquad \text{and} \qquad
\eta = \sum_{i=1}^n y_i \ind_{A_i},
$$
where $x_1\le x_2 \le \ldots \le x_n$, $y_1\le y_2 \le \ldots \le y_n$ and $\bigsqcup\limits_{i=1}^n A_i = \Omega$, $\Prob(A_i) = \frac1n$.
\\Let 
$$
\delta \defeq \sum_{i=1}^n \delta_i \ind_{A_i}, 
$$
where $\delta_1,\ldots,\delta_n \ge 0$ are defined iteratively by
\begin{equation} \label{eq:lem:def_delta}
\delta_k \defeq \max\left(
0,\; \sum_{i=1}^k y_i  - \sum_{i=1}^k x_i  - \sum_{i=1}^{k-1} \delta_i 
\right), \qquad k=1\ldots n.
\end{equation}
Such selection of $\delta_k$ instantly gives
\begin{equation} \label{eq:lem:almost_sosd}
\sum_{i=1}^k (x_i + \delta_i) \ge \sum_{i=1}^k y_i
\qquad \text{for all} \quad k=1\ldots n.
\end{equation}
Let us show that
\begin{equation} \label{eq:lem:ordered}
x_k + \delta_k \le x_{k+1} + \delta_{k+1} 
\qquad \text{for all} \quad k=1\ldots n-1.
\end{equation}
Indeed, if $\delta_k = 0$, than~\eqref{eq:lem:ordered} follows from $x_k \le x_{k+1}$ and $\delta_{k+1} \ge 0$.
If, however, $\delta_k > 0$, then
$\sum_{i=1}^k y_i  - \sum_{i=1}^k x_i  - \sum_{i=1}^{k} \delta_i = 0$, and hence, together with~\eqref{eq:lem:almost_sosd}, this gives
\begin{multline*}
x_k + \delta_k 
= \max\left(x_k, y_k + \sum_{i=1}^{k-1} y_i - \sum_{i=1}^{k-1} x_i - \sum_{i=1}^{k-1} \delta_i\right)
\le \max(x_k,y_k)
\le\\\le \max(x_{k+1}, y_{k+1})
= \max\left(x_{k+1}, y_{k+1} + \sum_{i=1}^k y_i - \sum_{i=1}^k x_i - \sum_{i=1}^k \delta_i\right)
= x_{k+1} + \delta_{k+1}.
\end{multline*}
Now, \eqref{eq:lem:almost_sosd} together with~\eqref{eq:lem:ordered} gives $\xi + \delta \sosd \eta$. 
\\Let $\gamma \defeq \gamma_n \cdot \ind_{A_n}$, where
$$
\gamma_n \defeq \sum_{i=1}^n y_i - \sum_{i=1}^n x_i - \sum_{i=1}^n \delta_i\ge 0.
$$
As a result, $\E (\xi + \delta) = \E (\eta + \gamma)$ and $\xi+\delta \sosd \xi+\gamma$.
By Lemma~\ref{lem:simple:sosd_implies_div1}, 
$\xi + \delta \div1 \eta + \gamma$.
Finally, by definition~\eqref{eq:lem:def_delta} of $\delta_k$,
\begin{multline*}
\sum_{i=1}^n \delta_i 
= \max\left(\sum_{i=1}^{n-1} \delta_i,\sum_{i=1}^n (y_i - x_i)\right)
=\\= \max\left(\sum_{i=1}^{n-2} \delta_i,
\sum_{i=1}^{n-1} (y_i - x_i), 
\sum_{i=1}^n (y_i - x_i)\right)=
\ldots 
= \max_{k=0\ldots n} \sum_{i=1}^k (y_i - x_i),
\end{multline*}
that essentially is~\eqref{lem:eq:mean_delta}.
\end{my_prf}

Now we are ready to prove the main result 
that complements Theorem~\ref{th:2sd:criteria}.
\begin{my_thm} \label{main:thm}
Let $\xi$ and $\eta$ be r.v.s with finite $\E \xi = \E \eta$.
Then the following statements are all equivalent:
\begin{itemize}
\item $\xi \sosd \eta$,
\item $\E{u(\xi)} \ge \E{u(\eta)}$ for all non-decreasing concave functions $u\colon \Real \to \Real$,
\item $\eta$ is a mean-preserving spread of $\xi$,
\item $\ES_\alpha ( \xi ) \le \ES_\alpha ( \eta )$ for all $\alpha \in (0,1]$,
\item $\xi {\divk1} \eta$,
\item $\xi {\divk2} \eta$.
\end{itemize}
In general case, when expectations not necessarily coincide (but still finite),
the second order stochastic dominance can be decomposed into 
the first order stochastic dominance and the diversification dominance, i.e.
if \ $\E|\xi|, \E|\eta|$ are finite, then $\xi \sosd \eta$ implies existence of $\zeta$, such that
$$
\xi \fosd \zeta \divk2 \eta,
$$
where $\xi \fosd \zeta$ means that $F_\xi(x) \le F_\zeta(x)$ for all $x\in\Real$.
\end{my_thm}

\begin{my_prf}
Clearly, $\xi {\divk1} \eta$ implies $\xi {\divk2} \eta$ by definition of closures and Lemma~\ref{lem:div_implies_sosd}, so 
we are going to prove two implications: 
$\xi {\divk2} \eta$ leads to $\ES_\alpha ( \xi ) \le \ES_\alpha ( \eta )$ 
for all $\alpha \in (0,1]$, 
which, in turn, leads to $\xi {\divk1} \eta$.
\\ 
First, suppose $\xi {\divk2} \eta$, i.e. there exist two sequences $\xi_n$ and $\eta_n$, such that $\xi_n \tok \xi$, $\eta_n \tok \eta$ and $\xi_n {\div2} \eta_n$ for all $n$. Due to Lemma~\ref{lem:div_implies_sosd} and Theorem~\ref{th:2sd:criteria}, the latter implies 
$$
\ES_\alpha (\xi_n) \le \ES_\alpha (\eta_n)
\qquad \text{for all} \quad \alpha \in(0,1] \text{ and } n \in \Natural.
$$
$\ES_\alpha$ is continuous w.r.t. Kantorovich metric:
$$
\left| \big. \ES_\alpha (\xi) - \ES_\alpha (\xi_n) \right|
\le \frac{1}{\alpha} \int_0^\alpha \left|q_\xi(u) - q_{\xi_n}(u) \right|\, du
\le \frac{1}{\alpha} \k(\xi,\xi_n) \to 0
$$
as $\xi_n \tok \xi$, and the same holds for $\eta_n$ and $\eta$, so
$$
\ES_\alpha(\xi) = \lim_{n\to\infty} \ES_\alpha(\xi_n) 
\le \lim_{n\to\infty} \ES_\alpha(\eta_n) = \ES_\alpha(\eta)
\qquad \text{for all} \quad \alpha \in (0,1].
$$
Now suppose $\ES_\alpha ( \xi ) \le \ES_\alpha ( \eta )$ 
for all $\alpha \in (0,1]$ and $\E\xi = \E\eta$.
It is known that the space of probability distributions on $(\Real,\sigmaB)$ with finite first moments endowed with Kantorovich metric is separable and complete
(see~\cite{Bolley2008}). The subset of all distributions, corresponding to simple r.v.s taking rational values with rational probabilities, is countable and everywhere dense. Therefore, there exist two sequences of simple r.v.s $\{\xi_n\}$, $\{\eta_n\}$, taking their values with rational probabilities, such that $\xi_n \tok \xi$ and $\eta_n \tok \eta$ as $n\to\infty$.
According to Lemma~\ref{lem:auxilary}, for every $n$ there exist non-negative r.v.s $\delta_n$, $\gamma_n$, such that
$\xi_n + \delta_n \div1 \eta_n + \gamma_n$ and
\begin{multline*}
0 \le \E \delta_n 
= \sup_{\alpha \in (0,1]} \left( 
    \int_0^\alpha q_{\eta_n}(u)\,du -\int_0^\alpha q_{\xi_n}(u)\,du \right)
\le\\\le 
    \sup_{\alpha \in (0,1]} \bigg( 
    \int_0^\alpha q_{\eta}(u)\,du 
    + \int_0^\alpha | q_{\eta_n}(u) - q_\eta(u) |\,du 
    \;- \\ \qquad\qquad\qquad
    - \int_0^\alpha q_{\xi}(u)\,du
    + \int_0^\alpha | q_{\xi_n}(u) - q_\xi(u) |\,du 
    \bigg)
\le\\\le
    \sup_{\alpha \in (0,1]} \alpha \cdot 
    \left(\big. \ES_\alpha(\xi) - \ES_\alpha(\eta) \right)
    + \k(\eta_n,\eta) + \k(\xi_n,\xi)
    \le \k(\eta_n,\eta) + \k(\xi_n,\xi)  \to 0,
\end{multline*}
so that $\E\delta_n \to 0$ as $n$ tends to infinity.
Hence, due to~\eqref{def:kantorovich_metric},
$$
\k(\xi_n+\delta_n, \xi_n) \le \E|\xi_n+\delta_n-\xi_n| = \E \delta_n \to 0
\quad \text{as } n\to\infty,
$$
and, therefore, by the triangle inequality,
$$
\k(\xi_n+\delta_n,\xi) \le \k(\xi_n+\delta_n,\xi_n) + \k(\xi_n,\xi) \to 0
\quad \text{as } n\to\infty.
$$
Next, $\E\gamma_n = \E{\xi_n} - \E{\eta_n} + \E\delta_n \to \E \xi - \E \eta = 0$ as $n\to\infty$, since the convergence in~Kantorovich metric implies the convergence of first moments.
Just as before,
$$
\k(\eta_n+\gamma_n,\eta) \le \k(\eta_n+\gamma_n,\eta_n) + \k(\eta_n,\eta) 
\le \E\gamma_n + \k(\eta_n,\eta) \to 0
\quad \text{as } n\to\infty.
$$
As a result, we constructed two sequences of r.v.s $\{\xi_n+\delta_n\}$ and
$\{\eta_n+\gamma_n\}$, such that $\xi_n+\delta_n \tok \xi$ and $\eta_n+\gamma_n \tok \eta$ as $n\to\infty$, while $\xi_n+\delta_n \div1 \eta_n+\gamma_n$.
We conclude that 
$$
\xi {\divk1} \eta.
$$
Let us prove the last statement of theorem.
Relation $\xi \sosd \eta$ implies $\E \xi \ge \E \eta$.
\\If~$\E \xi = \E \eta$, then one can take $\zeta \defeq \xi$.
If, however, $\E \xi > \E \eta$, then consider a function
$$
g(y) \defeq \E \min(\xi, y) = y - \E (\xi - y)^- 
= y - \int_{-\infty}^y F_\xi(x) \, dx.
$$
It has a non-negative derivative, and thus, 
$g$ is non-decreasing with range $(-\infty, \E\xi)$.
Hence, there exists $c\in\Real$, such that $g(c) = \E \eta$. 
Let $\zeta \defeq \min(\xi,c)$.
Clearly, $\xi {\fosd} \zeta$.
Finally, for every $a\in\Real$ one has
$$
\int_{-\infty}^a F_\zeta(x) \, dx 
= \int_{-\infty}^{\min(a,c)} F_\xi(x) \, dx + (a-c) \cdot \ind_{\{a>c\}}
\ge \int_{-\infty}^a F_\eta(x) \, dx,
$$
i.e. $\zeta \sosd \eta$, and by the first part of the theorem, 
$\zeta {\divk2} \eta$.
\end{my_prf}

\begin{my_cor}
The closure of relation of diversification is a partial ordering 
on the set of distributions with fixed (finite) first moments.
\end{my_cor}

\bibliographystyle{ieeetr}
\bibliography{diversify}

\end{document}